\documentclass[usenatbib]{mn2e}

\usepackage{graphicx}
\usepackage[usenames]{color}
\usepackage{amssymb}
\usepackage{appendix}
\usepackage{rotating}
\usepackage{amsmath}

\def\HI{\rm HI}
\def\HII{\rm HII}
\def\HeI{\rm HeI}
\def\HeII{\rm HeII}
\def\HeIII{\rm HeIII}

\newcommand{\zz}{z}
\newcommand{\nH}{n_{\rm gas}}

\newcommand{\Mpch}{h^{-1} \mbox{Mpc}}
\newcommand{\msun}{\mbox{M}_{\odot}}
\newcommand{\dxi}{x_{\rm th}}

\newcommand{\cgas}{C_{\rm gas}}

\begin{document}

\date{}

\pagerange{\pageref{firstpage}--\pageref{lastpage}} \pubyear{}

\title[Clumping factors for helium and hydrogen]
{Clumping factors of HII, HeII and HeIII} \author[A. Jeeson-Daniel
  et al.]{Akila Jeeson-Daniel,$^{1,2,3}$\thanks{E-mail:
    ajeeson@unimelb.edu.au} Benedetta Ciardi$^{2}$ and Luca Graziani$^{2}$\\$^{1}$School of Physics, University of Melbourne, 
  Parkville, VIC 3010, Australia\\ $^2$Max Planck Institute
  for Astrophysics, Karl-Schwarzschild Stra\ss{}e 1, 85741 Garching,
  Germany\\ $^3$ARC Centre of Excellence for All-Sky Astrophysics (CAASTRO)}

\maketitle

\label{firstpage}

\begin{abstract}
Estimating the intergalactic medium ionization level of a region needs
proper treatment of the reionization process for a large
representative volume of the universe. The clumping factor, a
parameter which accounts for the effect of recombinations in unresolved, small-scale structures, aids in achieving
the required accuracy for the reionization history even in simulations
with low spatial resolution. 

In this paper, we study for the first time the redshift
evolution of clumping factors of different ionized species of H and He
in a small but very high resolution simulation of the
reionization process. 
We investigate the dependence of the value and redshift  evolution of clumping factors
on their definition, the ionization level of the gas, the grid
resolution, box size and mean dimensionless density of the simulations. 
\end{abstract}


\section{Introduction}
\label{sec:reionintro}

Simulating the reionization history is a complex task due to the wide
range of spatial and mass scales which must be considered
\citep[e.g.][and references therein]{ciardi05,barkana07,morales10}. Due to the patchy nature
of reionization, large comoving volumes ($\ge 100~\Mpch$) are required
to representatively sample the dark matter halo distribution \citep{barkana04} and
to contain the large ionized regions (typically several tens of Mpcs in
size) expected prior to overlap \citep{wyithe04}. Concurrently, high mass resolution
is also needed to resolve the low mass galaxies thought to dominate
the ionizing photon emission \cite[e.g.][]{bolton07}, as well as the Lyman-limit systems which
determine the ionizing photons' mean free paths during the final
stages of reionization \citep{mcquinn07}.  Incorporating these sources and sinks of
ionizing photons correctly into the simulations requires spatial
scales of 10 kpcs to be resolved \citep{schaye01,mcquinn11}.

Performing N-body simulations which include gas hydrodynamics and the
multi-frequency radiative transfer (RT) of ionizing photons in a volume
with the required spatial and mass resolution is therefore a Herculean
task, with computational limits dictating the resolution one can
achieve. To alleviate this problem, it is possible to simulate large
computational volumes at a limited resolution and instead employ
sub-grid prescriptions for the physics which would be otherwise
missed. A typical example in reionization simulations is the adoption
of a clumping factor, a quantity which accounts for the effect of
recombinations in unresolved, small-scale structures on calculations
of the intergalactic medium (IGM) ionization state \citep[e.g.][]{madau99,iliev07,mcquinn07,kohler07}. 

In recent years, a significant amount of effort has therefore gone
into calculating the clumping factor of gas during the epoch of
reionization
\citep[e.g][]{giroux96,gnedin97,iliev05,trac07,mcquinn07,kohler07,pawlik09,raivcevic10,emberson12,shull12,finlator12,so13,kaurov13}.
Early calculations (e.g.  \citealt{gnedin97}) found high values for
gas clumping factors ($C_{\rm gas}\sim 40$) approaching the end of
reionization.  However, more recent studies have demonstrated that there
are wide variations in clumping factor values depending on how the quantity is defined and on feedback effects.  \cite{pawlik09} have
demonstrated that photo-heating during reionization significantly
reduces the clumping factor of gas ($C_{\rm gas}\sim 3$ at $z=6$),
thus lowering the number of ionizing photons needed to balance
recombinations and hence keep the universe ionized.  In another study,
\cite{raivcevic10} noted that clumping factors depend sensitively on
the volume over which they are computed within a simulation.  Other
authors have also studied the impact of different physical quantities
on the determination of the clumping factor.  For example, both
\cite{pawlik09} and \cite{shull12} discuss the effect of selecting a
density range for computing the clumping factor, while
\cite{finlator12} showed the importance of adopting appropriate
temperature and ionization thresholds.

However,  previous studies have focused on computing the
clumping factor for either all the gas in the IGM or for ionized
hydrogen only, while they have generally ignored the clumping
factor of ionized helium.  This is because the majority of
large-scale reionization simulations do not follow the ionization of
intergalactic helium {\it in addition to} hydrogen \citep[but see
  e.g.][]{trac07,pawlik11}.  However, recent work by \cite{ciardi11}
has demonstrated that a treatment of both hydrogen and helium using
multi-frequency RT is essential for accurately
computing the IGM temperature and (to a lesser extent) the H
ionization structure during reionization.  Future large simulations
should therefore ideally follow both hydrogen and helium ionization,
and unless these simulations are able to resolve all relevant scales
they will need to assume appropriate clumping factors for helium as
well.  For this reason, in this work we present estimates of the
clumping factor for ionized helium {\it and} hydrogen, and explore
their redshift evolution simultaneously by using a suite of high
resolution, multi-frequency RT simulations.  We also
examine in detail the quantities affecting the determination of
these clumping factors, such as resolution, gas density
and its distribution.

This paper is organized as follows.  We begin in
Section~\ref{sec:csim} where we briefly describe our radiative
transfer simulations.  We then examine
the clumping factor and its definition in detail in
Section~\ref{sec:clumping}.  Finally, we present our conclusions in
Section~\ref{sec:summary}. The cosmological parameters used throughout
this paper are as follows: $\Omega_{0,m}$=0.26,
$\Omega_{0,\Lambda}=0.74$, $\Omega_{0,b}=0.024~h^{-2}$, $h$=0.72,
$n_{s}$=0.95 and $\sigma_{8}$=0.85, where the symbols have their usual
meaning.

\section{Simulations of Reionization}
\label{sec:csim}

\begin{table*}
  \centering
  \caption[Reionization simulation details]{The hydrodynamical
    simulations used in this work.  The columns list, from left to
    right, the simulation identifier, the comoving box size $L$, the
    total number of particles (DM and gas), the DM particle mass
    $m_{\rm DM}$, the gas particle mass $m_{\rm gas}$, the comoving
    softening length $\eta$, the sampling grid size
    N$_{\rm c}^{3}$,  the sampling cell comoving size $L_{\rm c}$ and with/without RT outputs.}
  \begin{tabular}{|c|c|c|c|c|c|c|c|c|c|c|c|}
    \hline
    Model   & $L~[h^{-1}\rm\, Mpc]$ & Particles & $m_{\rm DM}~[h^{-1}M_{\odot}]$ & $m_{\rm gas}~[h^{-1}M_{\odot}]$ & $\eta~[h^{-1}\rm\, kpc]$ & $N_{c}^{3}$ & $L_{c}~[h^{-1}\rm\, kpc]$ & RT\\
    \hline
    $8.8G128$  &  8.78  & $2\times 256^{3}$   & $1.03\times 10^{7}$ & $5.19\times 10^{5}$ & $1.14$ & $128^{3}$ & 68.75 & Y\\
    $4.4G128$  &  4.39  & $2\times 256^{3}$   & $1.29\times 10^{6}$ & $6.48\times 10^{4}$ & $0.57$ & $128^{3}$ & 34.38 & Y\\
    $4.4G64$  &  4.39  & $2\times 256^{3}$   & $1.29\times 10^{6}$ & $6.48\times 10^{4}$ & $0.57$ & $64^{3}$ & 68.75 & Y\\
    $2.2G448$  &  2.20  & $2\times 256^{3}$   & $1.61\times 10^{5}$ & $8.10\times 10^{3}$ & $0.29$ & $448^{3}$ & 4.91 & N\\
    $2.2G384$  &  2.20  & $2\times 256^{3}$   & $1.61\times 10^{5}$ & $8.10\times 10^{3}$ & $0.29$ & $384^{3}$ & 5.73 & N\\
    $2.2G256$  &  2.20  & $2\times 256^{3}$   & $1.61\times 10^{5}$ & $8.10\times 10^{3}$ & $0.29$ & $256^{3}$ & 8.59 & N\\
    $2.2G128$  &  2.20  & $2\times 256^{3}$   & $1.61\times 10^{5}$ & $8.10\times 10^{3}$ & $0.29$ & $128^{3}$ & 17.19 & Y\\
    $2.2G64$  &  2.20  & $2\times 256^{3}$   & $1.61\times 10^{5}$ & $8.10\times 10^{3}$ & $0.29$  & $64^{3}$ & 34.38 & Y\\
    $2.2G32$  &  2.20  & $2\times 256^{3}$   & $1.61\times 10^{5}$ & $8.10\times 10^{3}$ & $0.29$  & $32^{3}$ & 68.75 & Y\\   
 \hline
\end{tabular}
\label{table:sims}
\end{table*}

The reionization simulations used here are based on those
recently described in detail by \cite{ciardi11}.  In this work, we
shall therefore only summarize their main characteristics.  Our
reionization simulations are performed by post-processing high
resolution cosmological hydrodynamical simulations with the 3D RT grid based Monte Carlo code {\tt CRASH}
\citep{ciardi01,maselli03,maselli09,pierleoni09,partl11,graziani12}. The
hydrodynamical simulations are performed with the tree-smoothed
particle hydrodynamics code {\small GADGET-3}, which is an
updated version of the publicly available code {\small GADGET-2}
\citep{springel05}. Haloes are identified at each redshift in the
cosmological simulations using a friends-of-friends algorithm with a
linking length of 0.2.  The hydrodynamical simulation snapshots --
which are sampled at regular redshift intervals -- therefore provide
the initial IGM gas distribution and halo masses.  These quantities
are then transferred to a uniform Cartesian grid of $N_c^3$ cells as input for {\tt CRASH}. It should be noted that the hydrodynamic simulations do not include a self-consistent treatment of radiative feedback, which is shown to cause a pressure smoothing of the gas and a reduction of the clumping factor (see e.g. \citealt{pawlik09}). Instead, an approximation of the heating of the IGM due to reionization is simulated by including an instantaneous photoionization and reheating by a spatially uniform ionizing background
\citep{haardt01} assuming an optically thin IGM.
In this work, we use hydrodynamical simulations performed in boxes of
comoving size 8.8, 4.4 and $2.2~\Mpch$. The parameters of these
simulations are summarized in Table~\ref{table:sims}.  It should be noted
that, although the procedure to run the simulations is the same as the
one described in detail in \cite{ciardi11}, here the size of the boxes is smaller.

Following the regridding of the hydrodynamical simulation data, the RT is then calculated as a post-process using {\tt
  CRASH}, which self-consistently follows the evolution of the
hydrogen and helium ($92$ and $8$ per cent abundance by number,
respectively) ionization states, along with the gas temperature.  All
the simulation boxes are gridded to an $N_c^3=128^{3}$ RT
grid, which leads to a lower spatial resolution for the larger boxes.
To study the effect of varying spatial resolution for the same box
size, we therefore also simulate the $2.2 ~h^{-1}\rm\, Mpc$ box with
coarser grid sizes of $64^{3}$ ($2.2G64$) and $32^{3}$
($2.2G32$). To test whether our gridding resolves the gas density distribution, we have also gridded the hydrodynamical simulation data to $256^3$ ($2.2G256$), $384^3$ ($2.2G384$) and $448^3$ ($2.2G448$) grid sizes. 
Due to the time required to run CRASH on grid sizes above $128^3$, the highest resolution for which we have a full RT simulation (i.e. from $z \sim 15$ to $z \sim 2$) is $2.2G128$, while the others have been used for testing purposes only. 
Furthermore, to understand the effect of varying box size
for the same spatial resolution, we have performed a $64^{3}$ simulation
in the $4.4 ~h^{-1}\rm\, Mpc$ box, which has the same spatial
resolution as $2.2G32$ and $8.8G128$.

All the RT simulations start at $\zz = 15$ and are run
until $\zz=2.2$. The emission properties of the sources are derived
assuming that the total comoving hydrogen ionizing emissivity at each
redshift is given by equation 3 in \cite{ciardi11}, and that the emissivity
of each source is proportional to its gas mass. The advantage of this empirical approach to assigning the ionization rate is that it avoids assuming an escape fraction of ionizing
photons, an efficiency of star formation and a stellar initial mass
function, which are very uncertain quantities. Note, however, that we still
need to assume an ionizing spectrum for the sources. 

Although we expect an evolution in redshift, with a predominance of sources with
harder spectra at lower redshift \citep[e.g.][]{haardt12}, in this work, 
following \cite{ciardi11}, we
have instead taken a simpler approach and chosen a fixed power-law
spectrum for the sources at all redshifts with an extreme-UV index of
$\alpha=1.8$, which is typical of the hard spectra associated with
quasars \citep{telfer02}. Simulation performed with this power-law 
spectrum is consistent with constraints on the $\HI$ photoionization rate 
from Ly$\alpha$ forest at $\zz\sim 6$ \citep{bolton07} and the Thomson 
scattering optical depth \citep{komatsu11} in the $35~\Mpch$ box simulations
of \cite{ciardi11}. Their requirement was that 
\HI~reionization is largely completed by $z \sim 6$, consistent with
observations of the \HI~Gunn-Peterson trough \citep[e.g.][]{fan06,bolton07}, and \HeII~reionization is largely completed by $z \sim 3$. In our small volume of $2.2~\Mpch$, though, due to the presence of high density regions which lead to higher recombination rates, 
the volume averaged ionization level of $\HII$ is 0.91 at $\zz \sim 6$, and reaches $0.99$ only at $\zz \sim 4$. $\HeII$~reionization, instead, is at $65$ per cent volume averaged ionization level by $\zz=2.2$. It should be noted that, despite being arbitrary, this
choice assures that the requirements mentioned above are met in a large 
representative volume of the universe. 

A preliminary exploration of the effects of varying the power-law spectral 
index $\alpha$ shows that, as expected, its value can affect both the ionization history of the various species and their clumping factors. If the volume average emissivity remains constant, a value of $\alpha$ in the range $1-3$ results in very similar H reionization histories,  while the effect on the evolution of $\HeII$ and $\HeIII$ is stronger because of the variation on the number of ionizing photons above $24.6$ eV and $54.4$~eV (refer to \citealt{ciardi11} for more details). Therefore, for different $\alpha$, the clumping factors of $\HII$ remain very similar, while those for $\HeII$ and $\HeIII$ are more strongly affected. We defer to future work a more detailed discussion of the effect of different source populations on the clumping factors.

\begin{figure}
\centering
\includegraphics[width=85mm,height=75mm]{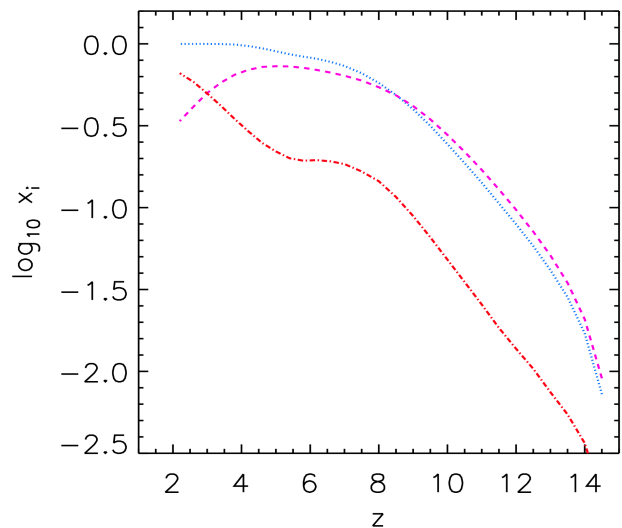}
\caption[Volume averaged ionization fraction of different species]{Redshift evolution of the volume averaged ionization fraction $x_{i}$ of different species in the $2.2G128$ simulation. The lines refer to: $i=\HII$ (blue dotted line), $\HeII$ (magenta dashed line) and $\HeIII$ (red dot dashed line).}
\label{fig:ionization}
\end{figure}

Here, we want to stress that the main aim of this paper is to study the evolution of the clumping factor of the various species and its dependence on a number of quantities, rather than properly model the reionization process (for which our simulation volumes are far too small). For this reason, we have maintained the same emission properties of the sources for all the simulations, while we plan to investigate the effects of the redshift evolution of emission properties in future work. For the purposes of this investigation then, the detailed reionization history of the simulations is not critical. Nevertheless, for the sake of clarity, in Fig.~\ref{fig:ionization} we plot the redshift evolution of the volume averaged ionization fractions of $\HII$, $\HeII$ and $\HeIII$ in $2.2G128$.
$x_{\HII}$ increases steadily to $\zz = 8$ and then slowly flattens out with the box reaching almost full ionization. A similar behaviour at $\zz >8$ is observed also in $x_{\HeII}$ and $x_{\HeIII}$, while the shape of the curves at lower redshift is determined by the shape of the emissivity (see also fig.~1 in \citealt{ciardi11}). At $\zz<6$ $x_{\HeII}$ ($x_{\HeIII}$) starts decreasing (increasing) as more $\HeII$ is converted to $\HeIII$. 

In the next section, we will investigate
the clumping factors and their dependences in more detail, using as a reference
simulation (unless otherwise noted) $2.2G128$. 

\section{Clumping Factor}
\label{sec:clumping}

In grid based simulations of reionization with the gas made of only H, the ionization
balance averaged over a cell can be written as
\citep[e.g.][]{kohler07}:

\begin{eqnarray}
\frac{\rm d}{\rm d \it t} \langle n_{\HI}\rangle_{\rm cell} = -3H\langle n_{\HI}\rangle_{\rm cell} -
C_{\rm I,\HI}\langle n_{\HI}\rangle_{\rm cell}\langle\dot{n}_{\gamma}\rangle_{\rm cell}
\nonumber\\ + C_{\rm R,\HII} \langle \alpha_{\rm R,HII}\rangle_{\rm cell}\langle
n_{\rm e}\rangle_{\rm cell}\langle n_{\HII}\rangle_{\rm cell}
\label{eqn:ckohler}
\end{eqnarray}

where $n_{\HI}$ and $n_{\HII}$ are the number density of neutral and
ionized hydrogen, $n_{\rm e}$ is the number density of electrons, $H$
is the Hubble parameter, $\dot{n}_{\gamma}$ is the ionizing photon
rate and $\alpha_{\rm R,\HII}$ is the recombination coefficient for
$\HII$. The angle brackets represent the mean value of the
distribution the cell volume would have, if the spatial resolution
were enough to resolve IGM structures down to the smallest relevant
scales. $C_{\rm I,\HI}= \langle n_{\HI}\dot{n}_{\gamma}\rangle_{\rm cell}/\langle
n_{\HI}\rangle_{\rm cell}\langle \dot{n}_{\gamma}\rangle_{\rm cell}$ is the clumping factor
of $\HI$ and $C_{\rm R,\HII}=\langle \alpha_{\rm R,\HII} n_{\rm e}
n_{\HII}\rangle_{\rm cell}/\langle \alpha_{\rm R,\HII}\rangle_{\rm cell}\langle n_{\rm
  e}\rangle_{\rm cell}\langle n_{\HII}\rangle_{\rm cell}$ is the clumping factor of
$\HII$. These can be used to estimate the ionization and recombination
rates in the cell due to unresolved small scale high density regions.

In this work, we want to calculate clumping factors of the IGM for the ionized species of H and He within our simulation box such that future work can use cell sizes equivalent to our box size while using clumping factors. Even though our best gridding resolution is poorer than the spatial resolution of the hydrodynamical simulation (see Section~\ref{sec:gridtests}), we aim to use our current simulations at limited resolution to calculate an estimate of He and H clumping factors and understand the degrees to which various factors, such as grid size, box size and overdensity of the region, would affect them.

Previous work have used a large number of equivalent definitions of clumping factors with different cuts in density ranges, ionization thresholds and with/without the effect of temperature on the recombination rates. To understand the effect of each of these modifications on the estimated clumping factor values, we start by starting from a simple definition of clumping factor and slowly progress to more complicated but accurate estimates of the true recombination number of each species.

As we are only interested in the clumping factor of the IGM, we need to
remove the cells containing collapsed haloes and large
dimensionless densities. We define the gas dimensionless density as
\begin{equation}
  \Delta =\nH/\langle \nH \rangle,
\label{eqn:deltadef}
\end{equation}
where $\nH$ is the gas number density in a cell and $\langle \nH
\rangle$ is the mean gas number density of the universe at that
redshift. The angle brackets $\langle\rangle$ give the value averaged within 
our box. This notation will be adopted throughout the paper, unless otherwise noted.
The dimensionless density of a collapsed dark matter halo depends on the
definition used to compute its virial radius. For example, for
spherical top hat collapse the dark matter dimensionless density at the virial radius is
$\sim 178$ \citep[e.g.][]{padmanabhan93}, while for an isothermal
collapse it is $\sim 60$ \citep[e.g.][]{lacey94}. For comparison and
consistency with earlier works, we adopt the generally used
dimensionless density threshold of 100, assuming that the IGM is composed only
of cells with $\Delta \leq 100$ \citep{escude00,pawlik09,raivcevic10}.

Our first definition of the clumping factor of a species is as follows:
\begin{equation}
C_{i}= \langle n_{i}^{2} \rangle/ \langle n_{i}\rangle^{2},
\label{eqn:cdef}
\end{equation}
where $n_{i}$ is the number density of the species $i$ =$\HII$, $\HeII$,
$\HeIII$~and total gas. This definition gives an estimate of the
uniformity in the distribution of each species by providing the spread
in the number density of species $i$ with respect to the mean value
calculated for the whole simulation volume. These clumping factors can be used e.g. in a one-zone model for reionization where the recombination term in the ionization balance equation is given by $C_{i} \langle n_{i}\rangle^{2}$.

Even though the above definition of $C_{\HII}$ is not very accurate considering that our gas conditions are different from the assumption in $C_{\rm R,\HII}$. Nevertheless, since most of the electrons come from H,  the $\HII$ clumping factor does not change substantially by including a more accurate evaluation of $n_{\rm e}$. The case B recombination rates vary only by a factor of a few for temperatures above $10^4$ K. We thus expect $C_{\HII} \sim C_{\rm R,\HII}$ also in our simulations. This holds true even for $\HeII$ and $\HeIII$, as the main source of electrons is mostly H (with electrons from He contributing mainly to the highly ionized regions up to about 17 percent in number) and the recombination rates of $\HeII$ and $\HeIII$ on temperature vary by only within about 4-5 times within the ionized regions in the IGM. Therefore, this definition provides a basic estimate of the clumping factors. The detailed effect due to variations in electron density and recombination rates on the clumping factors is explored in Section~\ref{sec:rec}.

\begin{figure}
\centering
\includegraphics[width=85mm,height=75mm]{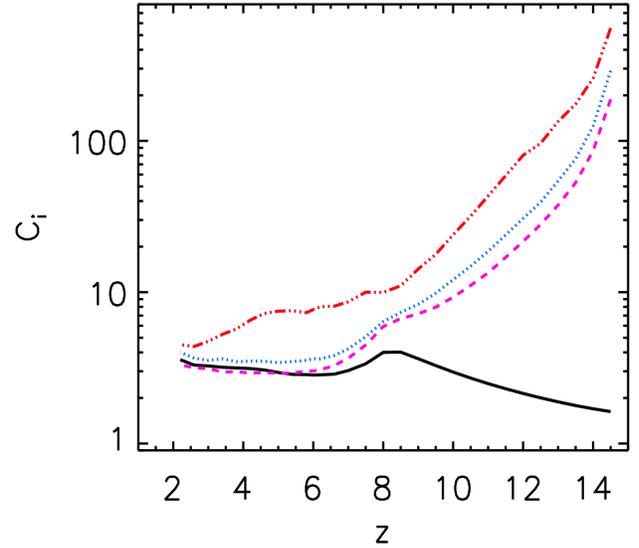}
\caption[Clumping factor $C$ of different species]{Clumping factor
  $C_i$ of different species in the $2.2G128$ simulation. The lines
  refer to: gas (black solid line), $\HII$ (blue dotted
  line), $\HeII$ (magenta dashed line) and $\HeIII$ (red triple dot dashed line).}
\label{fig:species}
\end{figure}
 
We start with studying how 
different species clump in our simulation $2.2G128$, by calculating the
redshift evolution of the clumping factor for total gas, $\HII$,
$\HeII$~and $\HeIII$, which are shown in Fig.~\ref{fig:species}. 
As previous works \citep[e.g.][]{pawlik09} typically compute the gas 
clumping factor, $C_{\rm gas}$, this is what we analyse first. 
Clumping factor for total gas, $C_{\rm gas}$, helps us understand 
the variation in the underlying gridded gas distribution in our simulation.
We find that $C_{\rm gas}$ increases
with decreasing redshift from 1.5 at $\zz \sim 15$ to about 3 at $\zz
\sim 8$. This trend is due to the self-gravity of the gas in the
IGM. At $\zz=9$ the hydrodynamical simulation includes instantaneous photoionization
and reheating of the IGM by a spatially uniform ionizing background
\citep{haardt01} assuming an optically thin IGM. This feedback leads
to pressure smoothing of the gas and to a reduction of the clumping
factor \citep[e.g.][]{pawlik09}, which decreases to about 2 at $\zz
\sim 6$. At $\zz\leq 6$ self-gravity becomes dominant again and the
gas clumping factor increases to 3 at $\zz \sim 2$. Note that the values of $C_{\rm gas}$ depend on the box size and the grid size of the simulation 
under study (in this case $2.2G128$), the effects of which will be discussed
in detail in Section~\ref{sec:tests}. 

Next we focus on the clumping factors of the ionized species $i=\HII,~\HeII$ and $\HeIII$, which take into account the effects of RT on the gridded gas distribution. In Fig.~\ref{fig:species}, $C_{\rm HII}$ reaches values as high as $\sim100$ at $\zz\sim14$, when only a handful of small ionized bubbles are present. Such values are obtained mainly because of the small fraction of ionized cells (i.e. with $n_{\rm HII}>0$) in the simulation volume and can be understood if one thinks about $C_{\rm HII}$ as representing the spread in the values of $n_{\rm HII}$ within the ionized cells (which is relatively large), divided by the fraction of ionized cells (which is very small).
When the reionization process is more advanced, lower $\HII$ clumping factor values are found. Eventually, $C_{\rm HII}$ converges to  $C_{\rm gas}$ at $\zz < 5$, when the volume averaged ionization fraction is $x_{\rm \HII} \gtrsim 0.95$. When the fraction of ionized cells approaches 1, i.e. all cells have $n_{\rm HII}>0$, $C_{\rm HII}$ represents the scatter in the $n_{\rm HII}$ values in the whole volume (which now is small because most cells are fully ionized). 

$C_{\rm HeII}$ closely follows the evolution of $C_{\HII}$, but with slightly lower values, 
due to the larger sizes of the $\HeII$ bubbles compared to the
corresponding $\HII$ regions.  This is due to a combination of the higher ionization cross-section of  $\HeI$ compared to that of $\HI$ \citep{osterbrock89} and the lower number density of He atoms (about 8.5 percent of the nuclei) in the gas compared to H, although this 
effect is partially balanced by the lower source photon rate at 24.6 eV compared to that at 13.6 eV ( $\sim 29$ percent). On the other hand, due to the low ionizing 
photon rate at 54.4 eV (for the assumed spectrum this is $\sim 8\%$ 
of the H ionizing photon rate) and high recombination rate (i.e. $\sim$ 5 times the one of $\HII$), $\HeIII$ is confined to small bubbles in the vicinity of the sources. Even at $\zz =2.2$, 
HeII reionization is not fully complete and only 65$\%$ of He is in $\HeIII$ state, 
while the rest is $\HeII$. Therefore, due to the highly spatially inhomogeneous 
distribution of $\HeIII$, $C_{\HeIII}$ has values higher than $C_{\HeII}$ and $C_{\HII}$, 
while the qualitative redshift evolution is similar for all the three clumping factors. We should note that, while we expect the same qualitative behaviour for different choices of the source spectrum, the quantitative results are bound to change. 

\subsection{Dependence on ionization level}

Earlier, we calculated the clumping factor $C_{i}$ including all cells, independently from their ionization level.  We saw that this leads to large values of the clumping factors at high redshift because of the small fraction of ionized cells and the relatively wide range in $n_{i}$. 
Typically, though, the clumping factor is employed to estimate the excess recombinations within the under resolved ionized volumes compared to those computed using the mean $n_{i}$ of such volumes. Therefore, removing the cells with $n_{i}=0$ by using ionization thresholds would give a better estimate of the clumping factors. For this reason, we also calculate the clumping factors of the different ionized species above a given ionization threshold $\dxi$, 
\begin{equation}
C_{i,\dxi}=\langle n_{i}^{2}\rangle_{x_{i}>\dxi}/ \langle n_{i}\rangle^{2}_{x_{i}>\dxi}
\end{equation}
where $x_{i}>\dxi$ denotes the set of cells with ionization of a species $i$ greater than $\dxi$. 
Since the number of recombinations depends on the square of $n_{i}$, the highest contribution to the clumping factors is given by the highly ionized, dense regions.

\begin{figure*}
\centering
\includegraphics[width=160mm,height=60mm]{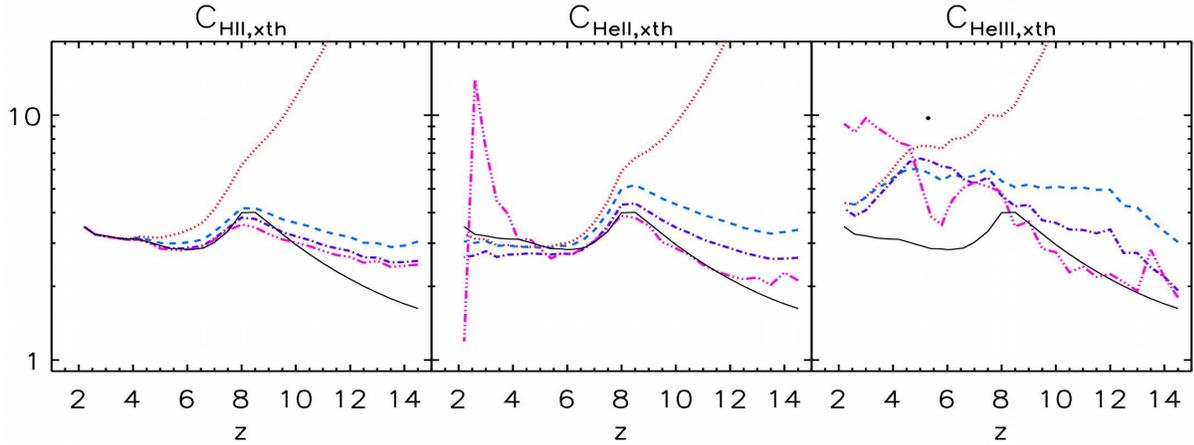}
\caption[Clumping factors in different ionization bins]{Clumping
  factors $C_{i,\dxi}$ against redshift for different ionization
  thresholds in the $2.2G128$ simulation. The different species
  plotted are (from left to right) $\HII$, $\HeII$ and $\HeIII$. The
  ionization thresholds are $\dxi=$ 0 (red dotted line), 0.1
  (blue dashed line), 0.5 (purple dot dashed line), 0.9 (magenta triple
  dot dashed line). The $C_{\rm gas}$ line is plotted in all the
  panels as a reference (solid black line).}
\label{fig:xion}
\end{figure*}

In Fig.~\ref{fig:xion}, we plot the redshift evolution of $C_{i,\dxi}$
for $i = \HII,~\HeII,~\HeIII$ and for $\dxi =~0,~0.1,~0.5,~0.9$,
together with $C_{\rm gas}$ for comparison. $C_{i,0}$ 
is the same as $C_{i}$ in Fig.~\ref{fig:species}.  All species
exhibit a decline in clumping factor values at high redshift when an ionization threshold $\dxi >0$ is introduced. This was expected as cells with $n_{i}=0$ are now removed from the calculation, inducing a reduction in the range of $n_{i}$ values.  Note that, in general, the probability distribution function of the selected $n_{i}$  becomes more peaked and less broad with increasing threshold values, leading to lower $C_{i,\dxi}$.

$C_{\HII,\dxi}$ for $\dxi >0$ is in the range
$2.5-4$ at $\zz>8$, with $C_{\HII,0.1}$ about $~0.25$ times larger than  $C_{\HII,0.5}$ and $C_{\HII,0.9}$. At $\zz<8$, the curves tend to
converge as most of the volume is fully ionized. $C_{\HeII,\dxi}$ has a trend 
similar to that of $C_{\HII,\dxi}$, with a larger difference (about 0.5) at $\zz>8$ 
between the curves for $\dxi>0$. At redshifts $\zz<6$, $C_{\HeII,0.1}$ and $C_{\HeII,0.5}$ are  lower than $C_{\rm gas}$ as these thresholds select a smaller range of cell densities. It is interesting to note that at $\zz <4$, $C_{\HeII,0.9}$ starts increasing because the number of cells with $x_{\HeII} >0.9$ decreases dramatically  as more and more $\HeII$ atoms are converted to $\HeIII$ close to the ionizing sources. At $\zz<3$, the curve is noisy, as less than 1 per cent of the cells have $x_{\HeII} >0.9$, with only 757 (18) cells satisfying the criterion at $\zz \sim 2.2~ (2.0)$. 

The curves for $\HeIII$ show similar values and trends, albeit noisier, as the reionization of
$\HeII$ to $\HeIII$ is a slow and inhomogeneous process which is dominated by the high
density regions around the sources, where two competing processes sculpt the reionization structure of $\HeIII$ - the ionization of the gas by sources and the fast recombination of $\HeIII$ to $\HeII$. There are very few cells with very high ionization. As a reference, less than 0.2 (10) percent of the cells have $x_{\HeIII}>0.9$ at $\zz=4.5$ (2.2), and even at $\zz=2.2$, 90 per cent of the cells have $0.1< x_{\HeIII}<0.9$. This leads to very noisy clumping factors, especially with high ionization thresholds. For $\dxi=0.5$ and $0.1$, the clumping factors have a more complex behaviour as the range of cells crossing these thresholds is larger. The clumping factors increase until $\zz =5$ and then they start decreasing as the ionization becomes more uniform, reducing the range of the selected $n_{\HeIII}$ values used to compute the clumping factors.

In general the trend is that clumping factors decrease with increasing ionization threshold. It should be kept in mind though that these clumping factors are calculated under the assumption that the recombination factor is constant, where as in reality it depends on the temperature. This dependence is explored in the next section.

\subsection{Dependence on temperature-dependent recombination coefficients}
\label{sec:rec}
\begin{figure*}
\centering
\includegraphics[width=160mm,height=60mm]{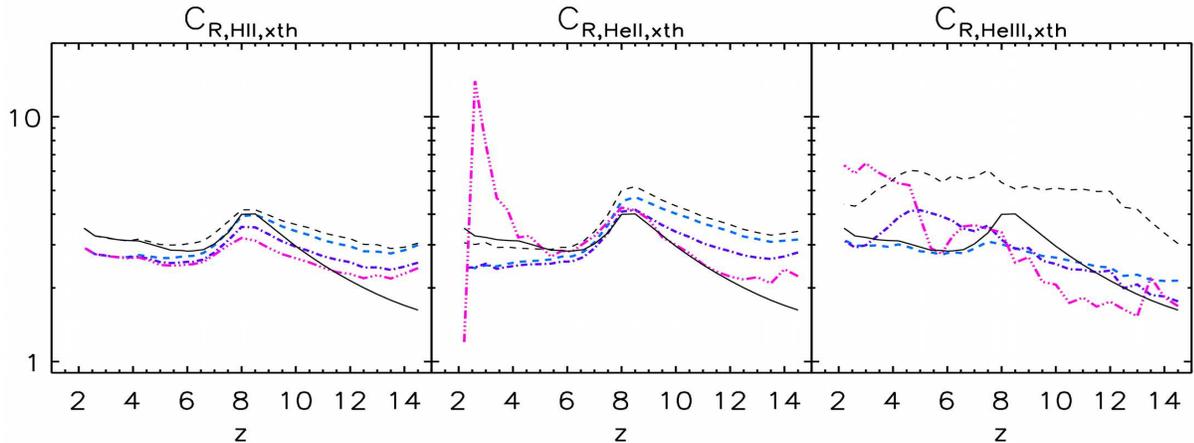}
\caption[Clumping factors in different ionization bins]{Clumping factors $C_{\rm R,\it i, \dxi}$ against redshift for different ionization thresholds $\dxi$ in the $2.2G128$ simulation. The species plotted are (from left to right) $\HII$, $\HeII$ and $\HeIII$. The ionization thresholds are 0.1 (blue dashed line), 0.5 (purple dot dashed line) , 0.9 (magenta triple ot dashed line). $C_{\rm gas}$ (black solid line) and $C_{\rm i, 0.1}$ (black dashed line) are plotted in all the panels as a reference.}
\label{fig:xionreem}
\end{figure*}

A better estimate of the clumping factors within an ionized volume is obtained by using the definition of \cite{kohler07}, as it takes into account also the information in the electron number density $n_{\rm e}$ and the temperature-dependent recombination rate $\alpha_{ \rm R, \it i}$ in each cell. In this case, the clumping factor, $C_{\rm R,\it i, \dxi}$,
is defined as:
\begin{equation}
C_{ \rm R, \it i, \dxi}=\langle \alpha_{\rm R, \it i} n_{\rm e} n_{\it i}\rangle_{x_{i}>\dxi}/\langle \alpha_{ \rm R, \it i}\rangle_{x_{i}>\dxi}\langle n_{\rm e}\rangle_{x_{i}>\dxi}\langle n_{\it i}\rangle_{x_{i}>\dxi}
\end{equation}
for different ionized species $\HII$, $\HeII$ and $\HeIII$, at ionization thresholds $\dxi =~0.1,~0.5,~0.9$.

Note that the effect of overdensity of the bubble regions in the ionization balance equation is also valid here.

Fig.~\ref{fig:xionreem} shows the redshift evolution of $C_{\rm R, \it i, \dxi}$ for the different ionization threshold $\dxi$ curves for $i=~\HII,~\HeII$ and $\HeIII$. Also shown in each panel are the curves for $C_{\rm gas}$ and $C_{i,0.1}$ for reference. Starting with $\HII$, it is easy to see that $C_{\rm R, \HII, \dxi}$ shows the same dependence of $C_{\HII,\dxi}$ on $\dxi$, i.e. the clumping factor decreases with increasing $\dxi$. $C_{\rm R, \HII, \dxi}$, though, has values lower than $C_{\HII,\dxi}$ due to highly ionized cells with higher temperatures, and thus lower recombination rates $\alpha_{ \rm R, \HII}$ than the mean in the volume. At low redshifts, this effect leads to a value of $C_{\rm R, \HII, \dxi}$ which is lower than both $C_{\HII,\dxi}$ and $C_{\rm gas}$ by about 0.5. 

$C_{\rm R, \HeII, \dxi}$ and $C_{\rm R, \HeIII, \dxi}$ show a behaviour very similar to that of $C_{\HeII, \dxi}$ and $C_{\HeIII, \dxi}$, respectively, with the former values being slightly lower due to the effect of temperature on $\alpha_{ \rm R, \HeII/\HeIII}$. This effect is stronger for $\HeIII$, as, for example, $C_{\rm R, \HeIII, 0.1}$ is almost half $C_{\HeIII, 0.1}$. 

Although the changes observed compared to $C_{i, \dxi}$ are mainly due to the effect of temperature-dependent recombination rates, also accounting for the contribution from both H and He to the electron number density affects the final results. The relative importance of a correct evaluation of $\alpha_{ \rm R, \it i}$ and $n_{\rm e}$ depends on the species considered and on redshift. While for $\HII$ and  $\HeII$ the effect of $\alpha_{ \rm R, \HII/\HeII}$ is dominant, for $\HeIII$ the changes induced by $n_{\rm e}$ are more relevant, although both $n_{\rm e}$ and $\alpha_{ \rm R, \HeIII}$ reduce the value of the clumping factor. As an illustrative example, on average, $C_{\HeIII, 0.1}$ is reduced by changes in $\alpha_{ \rm R, \HeIII}$ alone by about $22\%$ compared to $C_{\rm R, \HeIII, 0.1}$, while  changes in $n_{\rm e}$ alone induce reductions of about $33\%$. If they are considered together the reduction is $45\%$.

\subsection{Resolution Tests}
\label{sec:tests}

In this sub-section, we investigate how the behaviour of the clumping factor is affected by the box and grid size, by performing a number of resolution tests. 

The impact of grid and box size is due to two factors. The first one is the recombination of the species in the cells, i.e. the larger rate of recombinations in the high density cells of a simulation with a higher resolution.  The second is the source properties in a simulation volume of a given box size, i.e. larger boxes have higher mass sources which emit a higher number of ionizing photons into the surrounding IGM; this impacts the topology of reionization (refer to Appendix~\ref{appendix:ionsim_restest} for a detailed analysis), although the volume averaged emissivity is the same by construction for all simulations. Therefore, while using clumping factors in simulations to achieve better accuracy, we need to take into account both the box size and the grid size  of the volume used to compute them. 

\subsubsection{Dependence on Grid Size}
\label{sec:gridtests}

\begin{figure}
\centering
\includegraphics[width=70mm,height=210mm]{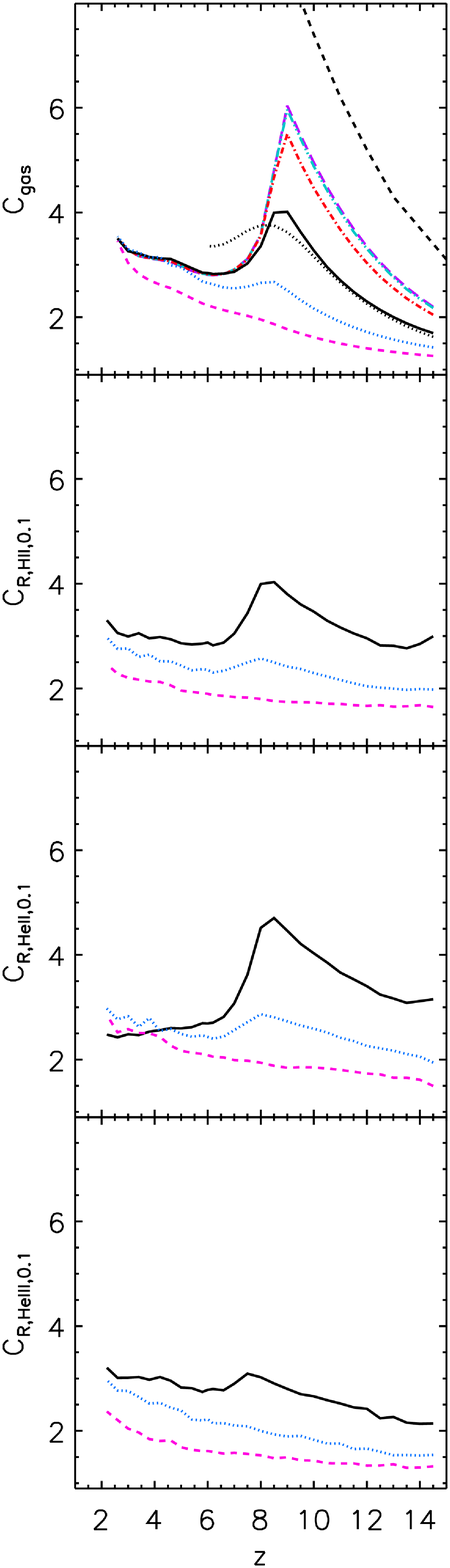}
\caption[Resolution test for grid size.]{Redshift evolution of the clumping factor $\cgas$ and  $C_{\rm R,\it i,\rm 0.1}$ for $i=\HII, ~\HeII$ and $\HeIII$ for the simulations $2.2G448$ (purple long dashed line), $2.2G384$ (cyan triple dot dashed line), $2.2G256$ (red dot dashed line), $2.2G128$ (black solid line; reference), $2.2G64$ (blue dotted line) and  $2.2G32$ (magenta dashed line). Also, plotted for comparison are the $C_{100}$ curve of $r9L6N256$ simulation of \cite{pawlik09} (black dotted line) and the $C_{100}$ for the $B2$ simulation of \cite{emberson12} (black dashed line).}
\label{fig:gridtest}
\end{figure}

Hydrodynamic simulations have a spatial resolution equal to the softening length of the simulation. On the other hand, the resolution of the RT calculation is determined by the sampling grid size, which we have limited to $128^{3}$ cells for our reference simulations. This leads to a lower spatial resolution in the RT calculations. To understand the effect this has on the clumping factor calculations, we compare simulations $2.2G448$, $2.2G384$, $2.2G256$, $2.2G128$, $2.2G64$ and $2.2G32$, which have the same hydrodynamic simulation outputs, but different sampling grid size. 

The top panel of Fig.~\ref{fig:gridtest} shows the evolution of $\cgas$ for $2.2G448$, $2.2G384$, $2.2G256$, $2.2G128$, $2.2G64$ and $2.2G32$. We also plot the $C_{100}$ curve of $r9L6N256$ simulation of \cite{pawlik09} and the $C_{100}$ for the $B2$ simulation of \cite{emberson12} for comparison.

We can see that our $\cgas$ values from the $2.2~\Mpch$ simulation converge at all redshifts for grid sizes above $384^3$ cells, while the gridded gas distribution is resolved at lower redshifts also with coarser grids. In our default simulation with grid size of $128^3$, $\cgas$ is resolved at $\zz<7$. 

Our $\cgas$ values from $2.2G128$  are comparable to the curve from \cite{pawlik09}. But even $\cgas$ from $2.2G384$ lies well below the estimate from \cite{emberson12}. This shows that even though a grid size of $384^3$ is enough to resolve the gas distribution within the simulation volume, the spatial resolution of the $2.2~\Mpch$ simulation with $256^3$ gas particles is not enough to capture the unheated IGM (Jeans mass $\sim~10^4 \msun$) at high redshifts. \cite{emberson12} find that a box size greater than 1 Mpc is needed, to sample the variance, along with a high mass resolution to resolve the $10^{4}~\msun$ gas clumps with at least $100$ particles. Our highest resolution simulation has a gas particle mass of $\sim 10^{4}~\msun$, explaining the lower values found for $\cgas$.
Once the IGM gas is heated to $10^4$ K, the Jeans mass becomes $\sim 10^{8}~\msun$ and is easily resolved in our $2.2~\Mpch$ simulation, although a direct comparison to \cite{emberson12} is not feasible as their simulations do not include heating of the gas due to reionization. The discrepancy observed at $\zz <9$ between our $\cgas$ and the value from \cite{pawlik09}  is due to the different strength of feedback effects which blow gas out from galaxies and to the low mass resolution of their simulation, as discussed in their paper.

In the bottom three panels of Fig.~\ref{fig:gridtest}, we plot the redshift evolution of $C_{\rm R, \it i, 0.1}$ for $i=\HII,~\HeII$ and $\HeIII$ for $2.2G128$, $2.2G64$ and $2.2G32$\footnote{Higher resolution grids have been run, but not until the lowest available redshift because of the long computation time. We find though that, as for the smaller grids, the behaviour of the clumping factor of the ionized species is similar to that of the gas.}. The three ionized species show a dependence on the grid resolution similar to that of $\cgas$, with higher clumping factors for increasing grid size. Differently from $\cgas$ and $C_{\rm R, \rm HII, 0.1}$ though, the clumping factors of helium do not seem to have reached convergence at low redshift. At high redshift a convergence would be reached only for grid sizes of at least $384^3$ (although the difference between a 256$^3$ and a 384$^3$ is minimal). 

From this test, we can see that even a $128^3$ grid does not resolve the small scale inhomogeneities present in the gas distribution. This means that either a clumping factor should be used for each cell (which is not the aim of this paper), or higher resolution RT simulations should be run\footnote{As already mentioned, simulations with grid resolution higher than $128^3$ have been run, but because of computation time they are not available for the full redshift range.}.  In this case we expect a slower reionization process (smaller ionized bubbles), due to the increment in the recombination rate from the better resolved high density regions. While both the gas and ionized species clumping factors increase with grid resolution, we expect the latter to be only slightly higher than the values shown in Fig.~\ref{fig:gridtest}, because smaller ionized regions produce lower values of clumping factors. This results in an increment milder than the one expected from an increase in gas clumping factors.

As mentioned in \cite{emberson12}, box size also plays an important role in determining clumping factor which is investigated in the next sub-section.

\subsubsection{Dependence on Box Size}

\begin{figure}
\centering
\includegraphics[width=70mm,height=210mm]{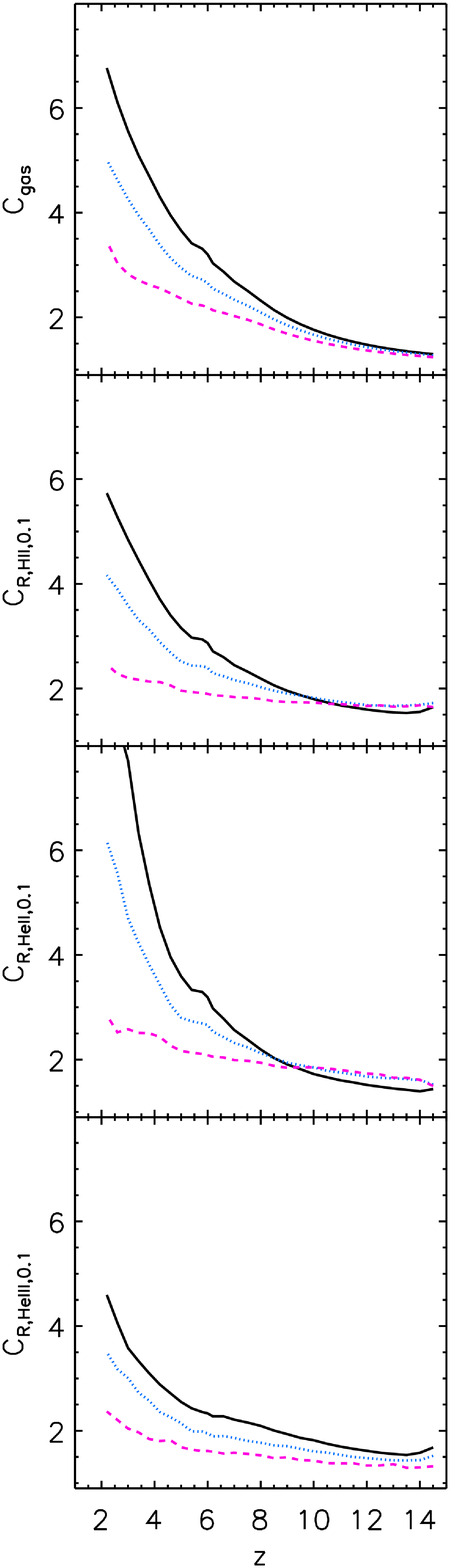}
\caption[Resolution test for box size.]{Redshift evolution of the clumping factor $\cgas$ and  $C_{\rm R,\it i,\rm 0.1}$ for $i=\HII, ~\HeII$ and $\HeIII$ for  simulations $2.2G32$ (magenta dashed line), $4.4G64$ (blue dotted line) and $8.8G128$ (black solid line).}
\label{fig:boxtest}
\end{figure}

Changing the box size while keeping the same spatial resolution affects the reionization process. Here we investigate the effect of the box size on the evaluation of the clumping factor using $2.2G32$, $4.4G64$ and $8.8G128$, which have the same sampling grid spatial resolution at different box sizes.

Fig.~\ref{fig:boxtest} plots the redshift evolution of $\cgas$ and $C_{\rm R, \it i, 0.1}$ for $i=\HII,~\HeII$ and $\HeIII$. We
can see that the box size affects $\cgas$ at $\zz < 9$. At
$\zz \sim 4$, e.g., $\cgas$ increases by $\sim 20$ per cent from
$2.2G32$ to $4.4G64$, and by the same amount to $8.8G128$. 
This is because in larger boxes at the same grid resolution, cosmic variance leads to cells with  dimensionless densities higher than in smaller boxes and thus to a larger range in gas density distributions, resulting in higher $\cgas$ values. As expected, the differences are more pronounced at lower redshift, when higher dimensionless densities are present. 

The clumping factors of all three ionized species ($i=\HII,~\HeII,~\HeIII$) show a qualitative behaviour similar to that of $\cgas$, i.e. clumping factors increasing with box size. As in the previous test, $C_{\rm R, \HeII, 0.1}$ shows the largest variation, with an increase in value of  up to $\sim 3$ between each box size step at $\zz=2.2$. The $\HeIII$ clumping factor shows a smaller change, with an increase of only about 1 in value between each box size step. Thus we can conclude that at a fixed spatial resolution, the box size does affect the estimate of the clumping factors, especially at low redshifts.

Other than resolution effects, the dimensionless density of the volume also seem to affect the calculations of clumping factors. This is studied in detail in the next sub-section.

\subsection{Dependence on Mean Gas Density}
\label{sec:subbox}

\begin{figure}
\centering
\includegraphics[width=65mm,height=210mm]{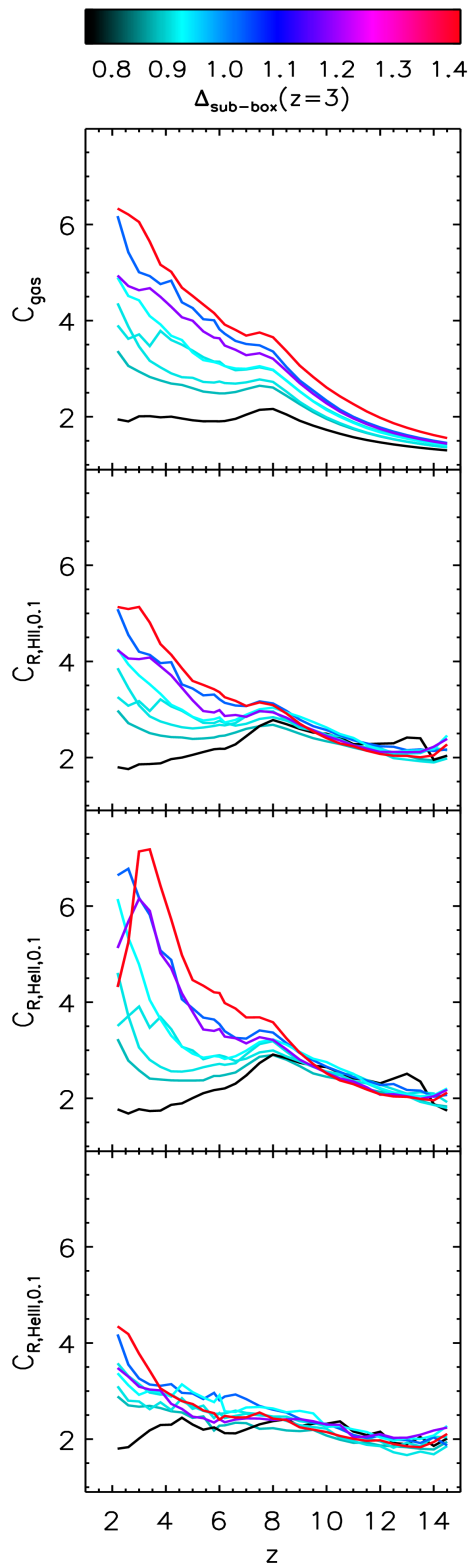}
\caption[Sub-box clumping factors.]{Evolution of the clumping factor $\cgas$ and $C_{\rm R, \it i, \dxi}$ for $i = \HII,\HeII$ and $\HeIII$ for ionization threshold $\dxi = 0.1$ in the eight sub-boxes of $64^3$ cells each from the $4.4G128$ simulation. The colours refer to $\Delta_{\rm sub-box}(\zz=3)$, which is the dimensionless density of the sub-box at $\zz=3$. The redshift evolution of the dimensionless densities of the sub-boxes has been shown in Fig. 8.}
\label{fig:subbox}
\end{figure}

Previous works have shown that the gas clumping factor correlates with
the gas density \citep[e.g.][]{kohler05,mcquinn07,kohler07}. \cite{raivcevic10} determined
that the sub-volumes within a simulation box have different gas clumping
factors due to differences in the gas distribution. To investigate this further, we split $4.4G128$ into eight sub-boxes with size $2.2~\Mpch$ of $64^3$ cells. Each of these
sub-boxes is equivalent to $2.2G64$, albeit with a different gas
distribution and source properties.

From the ionization history of the different
sub-boxes at the same grid resolution, we find that the higher the
mean density of the sub-box is, the higher is the probability of having large ionizing sources and the faster is the reionization. This is due to the clustering of high mass objects present in the high dimensionless density regions within that volume, resulting in a larger number
of ionizing photons into the IGM. Thus we can suspect that this would affect the clumping factor evolution. 

\begin{figure}
\centering
\includegraphics[width=75mm,height=85mm]{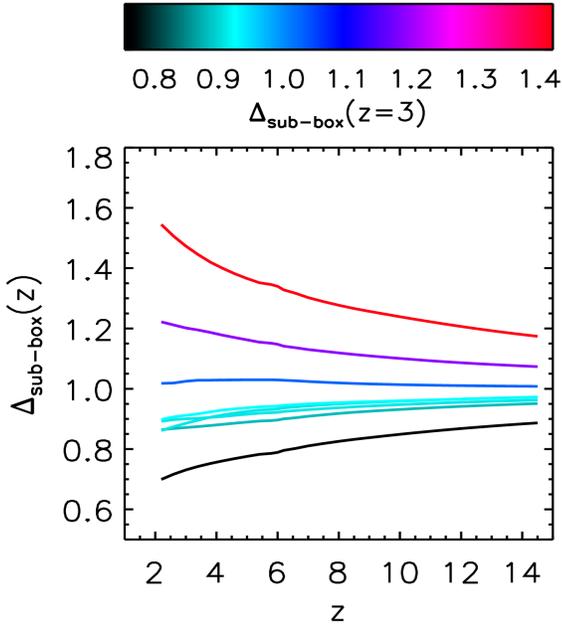}
\caption[Dimensionless density of sub-boxes.]{Evolution of the dimensionless density $\Delta_{\rm sub-box}$ in the 8 sub-boxes of $64^3$ cells each from the $4.4G128$ simulation. The colours refer to $\Delta_{\rm sub-box}(\zz=3)$, which is the dimensionless density of the sub-box at $\zz=3$. The $\Delta_{\rm sub-box}(\zz=3)$ values are (bottom to top) -  0.75, 0.88, 0.91, 0.91, 0.92, 1.03,  1.19, 1.42.}
\label{fig:subbox_overdens}
\end{figure}

Fig.~\ref{fig:subbox} shows the redshift evolution of $\cgas$ and
$C_{\rm R, \it i, 0.1}$ for $i =~\HII,~\HeII,~\HeIII$ in the eight sub-volumes. The lines for the
different sub-boxes are coloured according to $\Delta_{\rm sub-box}(\zz=3)$, i.e. the dimensionless density of the sub-box at $\zz=3$. While the qualitative behaviour of the clumping factor-dimensionless density correlation does not depend on our choice of the reference dimensionless density, the quantitative results do. It should be noted that the dimensionless density of a region is expected to vary slightly (by $20-30$ per cent between $\zz=15$ and 3) with redshift due to the flow of gas within neighbouring sub-boxes. For the sake of clarity, the redshift evolution of the dimensionless density of the sub-boxes is shown in Fig.~\ref{fig:subbox_overdens}.

In the first panel of the figure, we see that $\cgas$ shows a large range of 
values (a factor of 3-10) for the different sub-boxes at each redshift. 
Sub-boxes with higher dimensionless densities have a larger fraction of cells with 
high gas density, leading to larger gas clumping factors. The range in gas
clumping factors is increasing with decreasing redshift. Note that there is a large scatter in clumping factors in sub-boxes with similar dimensionless density. This is the probable reason for the high value of clumping factors for the sub-box with dimensionless density of $\Delta_{\rm sub-box}(\zz=3) = 1.03$ compared to the $\Delta_{\rm sub-box}(\zz=3) = 1.19$ case.

The three ionized species $i=\HII, \HeII, \HeIII$ exhibit very similar clumping factors at $\zz>10$, while differences emerge towards lower redshift, when a dependence on the dimensionless density of the region emerges. $\HII$ has a trend similar to that of the gas, with high dimensionless density regions showing high clumping factor
values. $\HeII$ exhibits a trend similar to that of $\HII$ at $\zz>4$, but once $\HeII$ gets converted to $\HeIII$ (which happens faster for high density gas), the regions in which $\HeII$ remains dominant have a smaller range of densities, resulting in lower clumping factors at low redshifts. In low density regions, instead, the conversion of $\HeII$ to
$\HeIII$ is slower because such areas are located further away from the sources
of ionizing radiation.

The reionization within low density regions is not dominated by
local sources, but rather it is influenced by the neighbouring
sub-regions. This leads to a large scatter, as can be seen in the
complex behaviour of $C_{\rm R, \HeIII, 0.1}$, which shows a general
trend of mildly higher clumping factors in high density regions except
for in the redshift range $4<\zz<8$, when some of
the cells in the low density regions start converting from $\HeII$ to
$\HeIII$, leading to high clumping factors. They then reduce as more
regions get reionized. These effects are stronger as we move to higher
ionization thresholds, i.e.  $\dxi=0.5,0.9$, but the curves are noisier due to
the small number of cells above such thresholds. 

As spatial resolution of $4.4G128$ is much below our best one $2.2G128$, we do not quantify the trends seen in these plots but qualitatively these are the trends we expect in general.

\section{Conclusions}
\label{sec:summary}

Simulating the ionization history of a representative volume of the
universe needs large box sizes, to encompass the patchy nature of the epoch of reionization 
and the massive ionized bubbles created towards the end of the
process. High spatial/density resolution is also necessary to
resolve the high-density Lyman-limit systems which control the 
evolution of reionization during its later stages. Since
simulating large comoving volumes with very high resolution is
computationally expensive, the general approach is to simulate large volumes at
a lower resolution, including sub-scale clumping factors to evaluate the effect of unresolved high density regions.

Recent work \citep[][]{ciardi11} has shown that He along with H plays an important role in determining the temperature and ionization structure of the IGM. In this work we employ similar simulations to study the different
factors affecting the reionization history of the IGM and the
estimation of clumping factors. We analyse a
suite of simulations of box and grid sizes in the range $2.2-8.8~\Mpch$ comoving and $32^{3}-448^{3}$, respectively.  

Using $2.2G128$, we calculate the clumping factor of the IGM $C_{i}$ for $i=$ gas, $\HII,\HeII, \HeIII$. $C_{\rm gas}$ has values in the range $1.5-3$ as in \cite{pawlik09}. The ionized species $\HII$ and $\HeII$ converge to the values of $C_{\rm gas}$ at $\zz < 6$, while at higher redshift they reach values as high as $\sim$ 100.  This is due to the inhomogeneous distribution of ionized gas in the simulation volume. $C_{\HeIII}$ has a qualitative behaviour similar to that of the other ionized species, but has higher values, due to the larger patchiness in the distribution of $\HeIII$.
  
The extremely high values of the clumping factor mentioned above are obtained because neutral cells are also included in the calculations. When only cells above a given ionization threshold, $\dxi$, are considered, the clumping factor $C_{i,\dxi}$ of the ionized species $i=~\HII,~\HeII,~\HeIII$ decreases to values closer to $C_{\rm gas}$. The above is true with the exception of $C_{\HeII,0.9}$, which increases at  $\zz < 5$, when $\HeII$ starts to be converted  to $\HeIII$ in high density regions. 
    
Finally, the more accurate definition of clumping factor $C_{\rm R, \it i,\dxi}$ is investigated, which shows that, due to the variation in the recombination coefficient with temperature and the correlation of electron density with ionization state, $C_{\rm R, \it i, \dxi}$ have slightly lower values than $C_{i,\dxi}$. The difference is the largest for $\HeIII$ clumping factors.

Grid size resolution tests on $\cgas$ show that to resolve the gas density distribution (and thus $\cgas$) over the entire redshift range in our $2.2~\Mpch$ box simulation with $256^3$ gas particles, we need a sampling grid of at least $384^3$ cells. $\cgas$ is instead already converged at $\zz<9$ in $2.2G128$. We find a general trend for $\cgas$ as well as for the ionized species, i.e. the clumping factors increase with increasing grid size. The evaluation of the clumping factors is also affected by the box size. We find that increasing the box size, while keeping the spatial resolution fixed, provides higher clumping factors.

The mean dimensionless density of the simulation volume plays a role in the determination of clumping factors as important as that of resolution effects. In most cases, clumping factors show a positive correlation with the mean dimensionless density, except for $\HeII$ during the later stages of reionization when it starts converting to $\HeIII$ in high density regions. This process induces a fast decrement of the clumping factor within such regions. 

\section*{Acknowledgements}

Many thanks to the anonymous referee for helpful comments. We thank James Bolton for providing us with the hydrodynamical simulations and for the constructive comments and suggestions throughout this work, Stuart Wyithe for the useful comments on the draft and Andreas Pawlik and Kristian Finlator for the stimulating discussions. AJD acknowledges support from and participation in the International Max Planck Research School for Astrophysics at the Ludwig-Maximilians University. Parts of this research were conducted by the Australian Research Council Centre of Excellence for All-sky Astrophysics (CAASTRO), through project number CE110001020. LG acknowledges the support of  DFG Priority Program 1573.

\appendix

\section{Ionization History for Resolution Test Simulations}
\label{appendix:ionsim_restest}

\begin{figure}
\centering
\includegraphics[width=80mm,height=180mm]{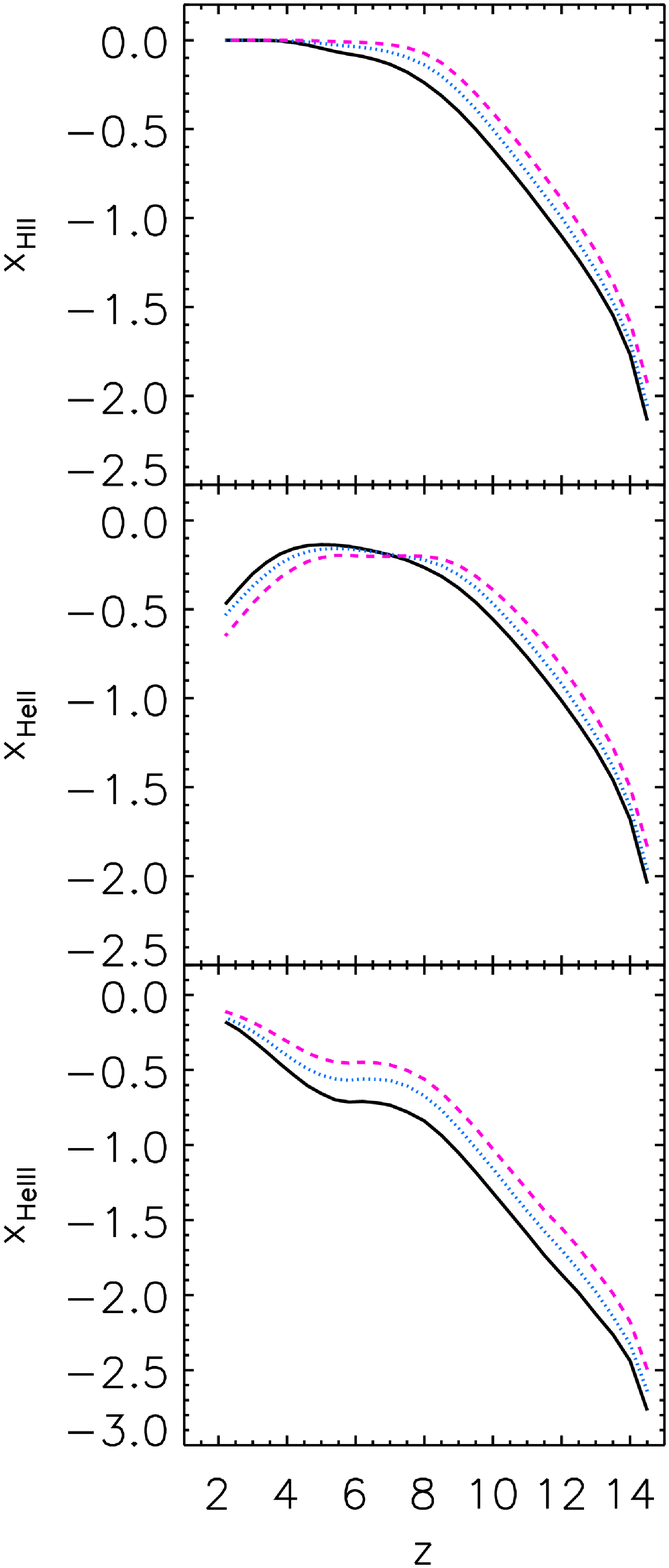}
\caption[Ionization fractions for grid size test simulations.]{Redshift evolution of the ionization fractions $x_{i}$ for $i=\HII, ~\HeII$ and $\HeIII$ for the simulations $2.2G128$ (black solid line; reference), $2.2G64$ (blue dotted line) and  $2.2G32$ (magenta dashed line).}
\label{fig:ionsim_gridtest}
\end{figure}

In this section, for the sake of clarity, we show how the grid and box size affect the ionization history. Fig.~\ref{fig:ionsim_gridtest} shows the fraction of $\HII$, $\HeII$ and $\HeIII$ for $2.2G128$, $2.2G64$ and $2.2G32$. All simulations have the same hydrodynamic resolution but different grid size. The general trend is that of a faster reionization of the volume with a decreasing grid size, due to the lower rate of recombination in low resolution simulations, showing the necessity of using a clumping factor in simulations which under-resolve the gas density distribution. This trend is inverted for $x_{\HeII}$ at $\zz<8$, where the conversion of $\HeII$ to $\HeIII$ dominates over the single ionization of $\HeI$.

\begin{figure}
\centering
\includegraphics[width=80mm,height=180mm]{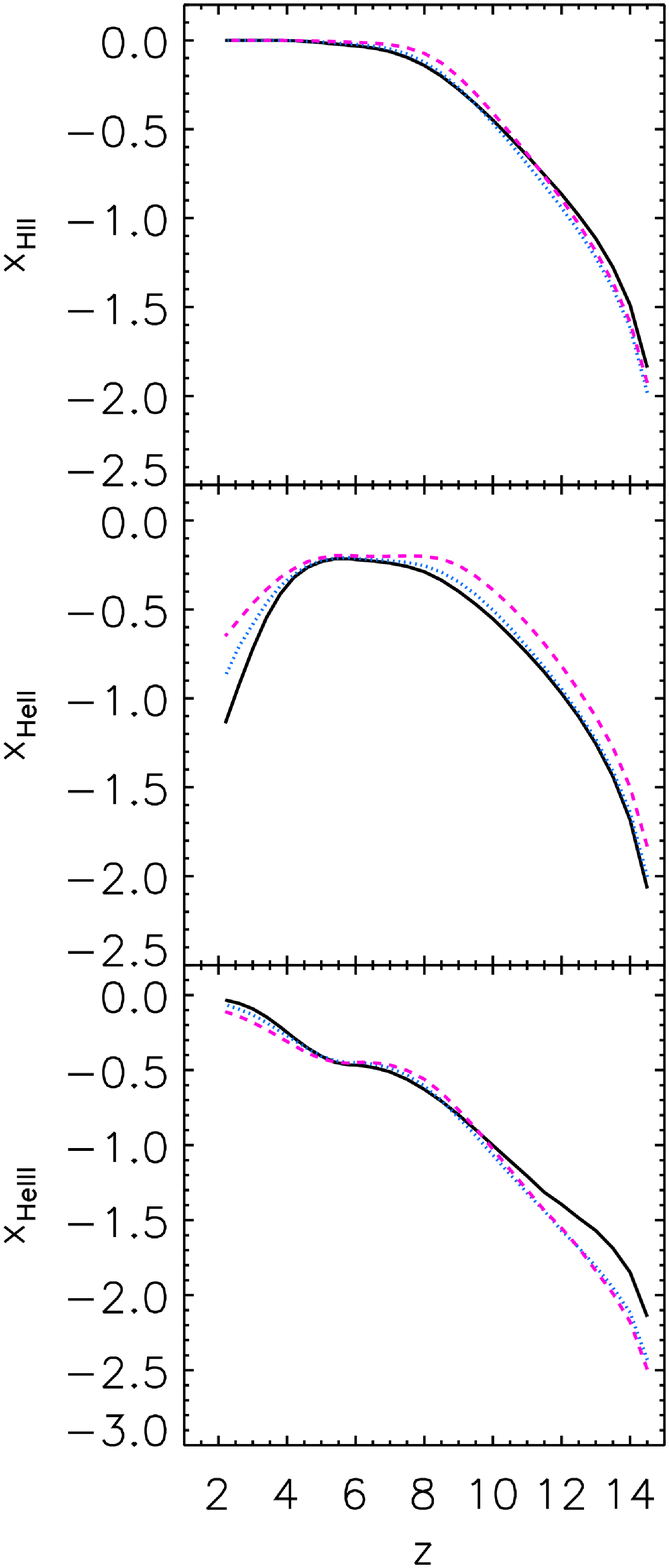}
\caption[Ionization fraction for box size test simulation.]{Redshift evolution of the ionization fractions $x_{i}$ for $i=\HII, ~\HeII$ and $\HeIII$ for  simulations $2.2G32$ (magenta dashed line), $4.4G64$ (blue dotted line) and $8.8G128$ (black solid line).}
\label{fig:ionsim_boxtest}
\end{figure}

In Fig.~\ref{fig:ionsim_boxtest}, we plot the fraction of $\HII$, $\HeII$ and $\HeIII$ for $2.2G32$, $4.4G64$ and $8.8G128$. These simulations have equal gridding resolution but different box sizes. For $\HII$ and $\HeIII$, the presence of stronger sources, due to cosmic variance, lead to higher ionization values in larger boxes at $\zz >12$. As the density field evolves at lower redshifts ($6<\zz<12$), the effect of stronger ionizing sources is mitigated by the contribution of higher recombination rates in the high density cells around these sources. Eventually at $\zz<6$, a strong rise in emissivity causes the ionization fractions in large boxes to increase rapidly, especially for $\HeIII$. In the case of $x_{\HeII}$, the high conversion rate of $\HeII$ to $\HeIII$ in large boxes leads to low $x_{\HeII}$ at all redshifts. But as the conversion of $\HeIII$ to $\HeII$ increase at intermediate redshifts ($6<\zz<12$), the recombination of $\HeII$ also increase causing the trend to stay the same till $\zz<6$, after which the high emissivity causes the ionization of $\HeII$ to become the dominant factor and thus reversing the trend.
It should be noted that in general the effect of box size is stronger for the He component of the gas, while the H is only mildly affected.

Therefore, we conclude that both grid size and box size play an important role in determining the ionization history of a volume which in turn affect the clumping factor values.

\label{lastpage}

\end{document}